\documentclass[11pt,twoside]{./atmp}

\usepackage{amsmath,amssymb}

\usepackage[all]{xy}
\usepackage{color}
\usepackage{graphicx}

\begin{document}

\title[Self-organization in the tornado: new approach]
{Self-organization in the tornado: the new approach in the tornado description}

%\arxurl{http://arxiv.org/}

\author[Bystrai G.P., Lykov I.A.]{Bystrai Gennadyi Pavlovich, Lykov Ivan Aleksandrovich}

\address{Ural Federal University\\51, Lenin avenue, Yekaterinburg, Russia, 620000}
\addressemail{Gennadyi.Bystrai@usu.ru}

\begin{abstract}
For the mathematical modeling of highly non-equilibrium and nonlinear processes in a tornado in this paper a new approach based on nonlinear equations of momentum transfer with function of sources and sinks is suggested. In constructing the model thermodynamic description is used, which is not entered before and allows discovering new principles of self-organization in a tornado. This approach gives fairly consistent physical results. This is an attempt to answer some fundamental questions concerning the existence of a tornado based on the created model and numerical results.
\end{abstract}

\maketitle
\cutpage
\noindent

\section{Introduction}

Tornado (Spanish Tornado ''Whirlwind''), twister -- rapidly rotating atmospheric vortex that occurs in cumulonimbus (thunderstorm) cloud and spreads down to the ground, in the form of cloud arms or trunk of tens or hundreds of meters.

This phenomenon exists for millions of years and it is also observed on other planets with atmosphere. But until now it is a little studied and hard to be measured owing to the presence of large air velocities near the funnel (kernel, trunk) and, consequently, owing to the inability of the measuring equipment to withstand powerful external influences\cite{Brooks}.

From a purely dynamical point of view, and tornadoes occur due to the strengthening of existing localized vortices by an external flow of momentum\cite{DavJ}. Supercell thunderstorm models\cite{Arsen,Brow,HF,LemD,Rasm,Schl} presented in the last forty years, have many similarities in their basic structure. However, in contrast to the purely dynamic approach to work, beginning with article\cite{Schl} all models have intensive upstream, which cross the vertical wind shear, increases its speed and swirl with height. Thus, these models are the convective and they are based on the model of mesocyclone, which formation has been just recognized as the most useful fact to supercell identification\cite{DavJ}. At the same time, they differ in the detail level from the occurring processes and as a consequence in the initial and boundary conditions that are necessary for the thunderstorm supercell occurrence. Numerical models of supercell thunderstorms\cite{KlWil,WicW} are based on the dominant mesocyclone. The hydrodynamic model\cite{WicW}, which takes into account a set of cross-produced pair and other interactions and forces, is particularly interesting.

Observations of the tornado have a rich history, provided by many papers only for the 20th century. Brooks was the first observer, who put forward generally accepted assumption, that the funnel is a part of the parent cloud, the structure and dynamics of which represent a small tropical storm and having a helical structure\cite{Brooks}. Numerous observations of the parent cloud indicated the presence of long vortices in the horizontal plane in them; the vertical poles (funnels) are the continuation of which\cite{Weg}. This fact has no explanation. In 1951 in Texas during a tornado the funnel, passing over the observer, rose, and its edge was at the height of 6 meters with the inner cavity diameter of 130 meters. The wall thickness was the size of 3 meters. Vacuum in the cavity was absent, because it was easy to breathe during its passage. The walls was extremely fast spinning (Justice, 1930). In the monograph Flora\cite{Flora} notes that "the distinction between the strongest winds in the body of the funnel and the stationary air on its periphery is so sharp, that it causes a number of damaging effects."

Observations of the actual tornado, therefore, indicate a strong non-linearity and non-equilibrium of processes in atmosphere during the formation and existence of a tornado, that does not allow to create the perfect model of this exotic phenomenon.

In the framework of the study of this unusual natural phenomenon the following questions are need to be answered:

1. Under what conditions in the atmosphere the appearance of a tornado happens?

2. What causes the existence of distinct lateral boundaries of the tornado? Why don't these boundaries spread in time?

3. What are the conditions for the existence of dissipative structures observed in the tornado -- a set of organized vortices? And what are their conditions of decay?

4. What causes the boundedness of tornadoes in height?

5. What are the conditions for the existence and stability of stationary and others possible modes.

6. What determines the appearance of the tornado core (trunk) with significantly higher velocities?

For the mathematical modeling of highly non-equilibrium and nonlinear processes in a tornado authors propose the approach based on the nonlinear equations of momentum transfer with the model sources and sinks function. This approach can be assigned to the problems with peaking considered by Academician A.~A.~Samarskii\cite{KKM,SamM,SKAM,SamC}. For the first time the thermodynamic description was used to identify new principles of self-organization in the atmosphere in the model specification not entered before.

\section{Model}

On the basis of the equations of momentum with a nonlinear function of sources and sinks the nonlinear hydrodynamic model of a strongly nonequilibrium processes in the atmosphere during an intense vortex formation was constructed by the authors. A thin layer of a unit volume, parallel to the surface of the earth, is allocated in the tornado. In this layer there are sources of the movement and its sinks depending on the horizontal velocity vector and its modulus. The problem is solved under the assumption of isothermality of the layer and its vicinity. Temperature and other air characteristics changes with the altitude. "Layered" model is used to simplify the modeling of vortices existing in the three-dimensional region of space, as well as to facilitate numerical calculations.

Non-linear momentum source in the atmosphere in some cases leads to the regime with peaking -- sharp increases in the velocity amplitudes during some short-time. The development of this mode leads to a self-organization. In this formulation, the formation of dissipative structures localized in a space, is possible. In these structures the speed may increase limited (or unlimited).

Authors using this model based on the thermodynamic approach, attempt to describe the observed physical phenomena and to explain the mechanisms of nonlinear layered momentum transfer in a tornado in case of the nonpotential flow.

In the non-equilibrium thermodynamics it is accepted to characterize the processes within the system under the influence of the external environment by the so-called entropy production $\sigma ^{i} $ per unit volume of the layer. There are also other local thermodynamic characteristics -- the external flow of entropy $\sigma ^{e} $ and the rate of change of entropy $\mathop{S}\limits^{\bullet } $, equal to their sum. It is believed that in the self-organization systems full change of the entropy decreases with time: $\mathop{S}\limits^{\bullet } <0$.

This approach allows one to record the system of equations for the velocity components in the case of an incompressible fluid in dimensionless form as two-dimensional Kuramoto-Tsuzuki equation\cite{KuTs} for the atmospheric layer:
\begin{equation}\
\frac{\partial \Phi }{\partial t} =\nu _{1}^{*} \left(1+ic_{1} \right)\left(\frac{\partial ^{2} \Phi }{\partial x^{*2} } +\frac{\partial ^{2} \Phi }{\partial y^{*2} } \right)+q^{*} \Phi -\alpha _{1}^{*} \left(1+ic_{2} \right)\left|\Phi \right|^{2} \Phi ,\
\end{equation}
where $\Phi =\vartheta _{x}^{*} +i\vartheta _{y}^{*} $; $A_{1} =\nu _{2}^{*} /\nu _{1}^{*} $ related to the viscosity, $A_{2} =\alpha _{2}^{*} /\alpha _{1}^{*} $ due to sinks. The superscript "*" means the dimensionless form of the parameters on certain scales, identified at the observing of a tornado.

\section{Results}
The results of numerical simulations give a good resemblance to the observed physical phenomena. Approach allows us to explain nonlinear mechanisms layered momentum transport in the atmosphere, and also origin, evolution and decay of large, medium and small atmospheric vortices as dissipative thermodynamic structures. Competition of pulse increment and propagation processes in viscous medium leads to the appearance of a linear size -- the spatial diameter of self-organized structure -- $l_{0}$ on the border of which the rate of change of entropy changes sign. You can set the dependence of the initial conditions from humidity, which varies with altitude.

\begin{figure}[h]
\center{\includegraphics[width=83.6mm, height=81.5mm]{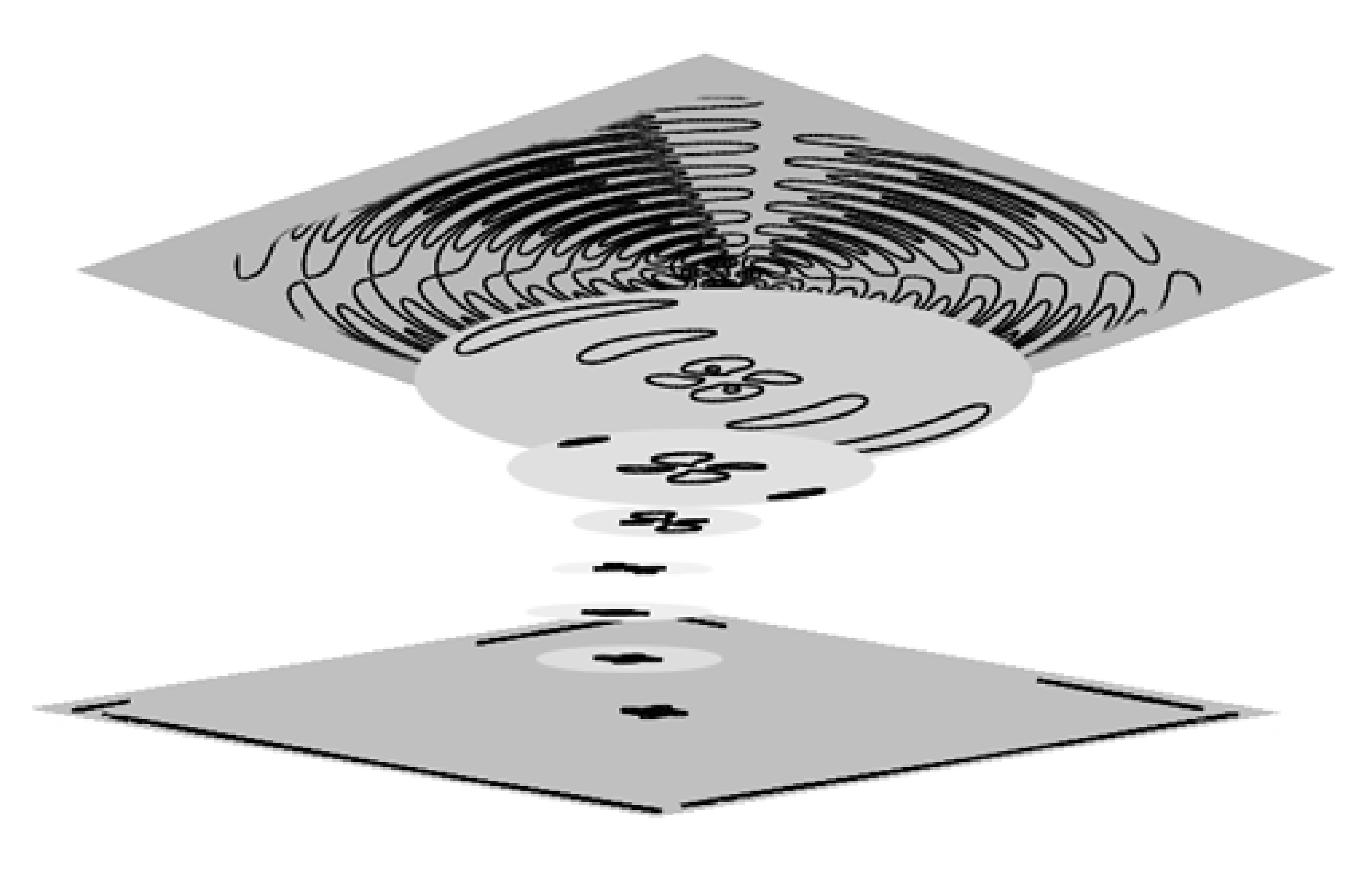}}
\caption{The entropy production in the tornado basin in layers at different heights from the ground up to 3.5 km.}
\label{fig1ent}
\end{figure}

Humidity, which determines the radial component of the emerging spiral wave, in the lower layers at its reduction leads to a significant depletion of the vortex structure around the trunk and even their disappearance, and, thus, to more realistic result when the trunk is only visible between the earth and cloud formed the tornado (Fig. \ref{fig1ent}).

The Table \ref{SpatTorn} shows the spatial characteristics of the comparatively small size tornado defined by numerical methods for two variants of the approximation of turbulent flows: nonpotential (NP) and potential (P). Nonpotential flow means that the function of the pressure gradient relates to the velocity in terms of the sources and sinks. Determination was made on the velocity ($\vartheta ^{*} $) and pressure gradient ($\nabla p^{*} $). The obtained characteristics are in a good agreement with observations of real tornado, that is according to the authors confirm the conformity between theory and experiment.

\begin{table}[hb]
\caption{\label{SpatTorn}Spatial characteristics of tornado with a diameter pool $l_{0}=500~m$, $\left|m\right|=1$, the comparative table.}
\begin{tabular}{@{}llll}
\hline
Flow type& NP & NP & P\\
\hline
{\it Characteristics} & $\vartheta ^{*} $ & $\nabla p^{*} $ & $\nabla p^{*} $ \\
\hline
Self-organization zone diameter \textit{d}, \textit{m} & 439 & 459 & 426 \\
Inner core diameter,\textit{ m} & 11 & 9 & 9 \\
Outer core diameter,\textit{ m} & 20 & 20 & 21 \\
Width of the pressure equalization \\
ring to atmospheric pressure, \textit{m} & 30 & 20 & 37 \\
\hline
\end{tabular}
\end{table}

Thus, the thermodynamic model is fundamentally different from the convective models, and allows you to construct model that reflects some of the principal features of the real observed tornado. This is such features as the presence of the trunk almost perfectly circular in cross section, the existence of significant gradients of velocity and pressure on the edge of the tornado basin and in the area of the vertical vortex tube\cite{Flora}, along with a complete absence of rotation and discharge in the center of the tornado\cite{Just}. The considered approach allows us to formulate conditions of self-organization in a tornado established by the authors in present. This approach allows us to expand the problem in the asymptotic to conditions of formation of cyclones in the atmosphere. Also some physical restrictions for the horizontal velocity, the intensity of the sources and sinks and the stability condition of dissipative structures are specified. These restrictions can not be obtained explicitly using the convective and other models.

Such a nonlinear hydrodynamic model based on a thermodynamic approach can be used to describe cyclones, tornadoes, and other exotic natural phenomena.

\subsection{Condition for the stability of vortex flows in a tornado.}

Stability analysis of solutions of the Kuramoto-Tsuzuki equation\cite{KuTs} yields the following result:

\[\left(c_{1}^{2} +1\right)k^{4} +2\left(1+c_{1} c_{2} \right)k^{2} >0, k=\pi /l_{0} .\]

In order to satisfy the inequality for any value of \textit{k} it is necessary to limit constants $c_{1} ,c_{2} $: $-1<c_{1} c_{2}< 1$. This condition is the criterion for the stability of turbulent structures in atmospheric vortices. Therefore, it is a condition of release of such vortices in the steady state.

The emergence of a stationary core in a plane layer of the tornado -- space localized figure -- is the result of sharpening and confirmed by the work\cite{Lane}, which describes the unusual properties of the presence of large pressure gradients and velocities in the tornado.

Real tornado in this case goes to steady state and has the core (trunk) formed in a planar layer, spatial characteristics of which are shown in Table~1.

As an intermediate generalization of this work we present currently installed thermodynamic and hydrodynamic conditions of self-organization.

\subsection{The thermodynamic conditions of self-organization.}

\begin{figure}[h]
\center{\includegraphics[width=120mm, height=62mm]{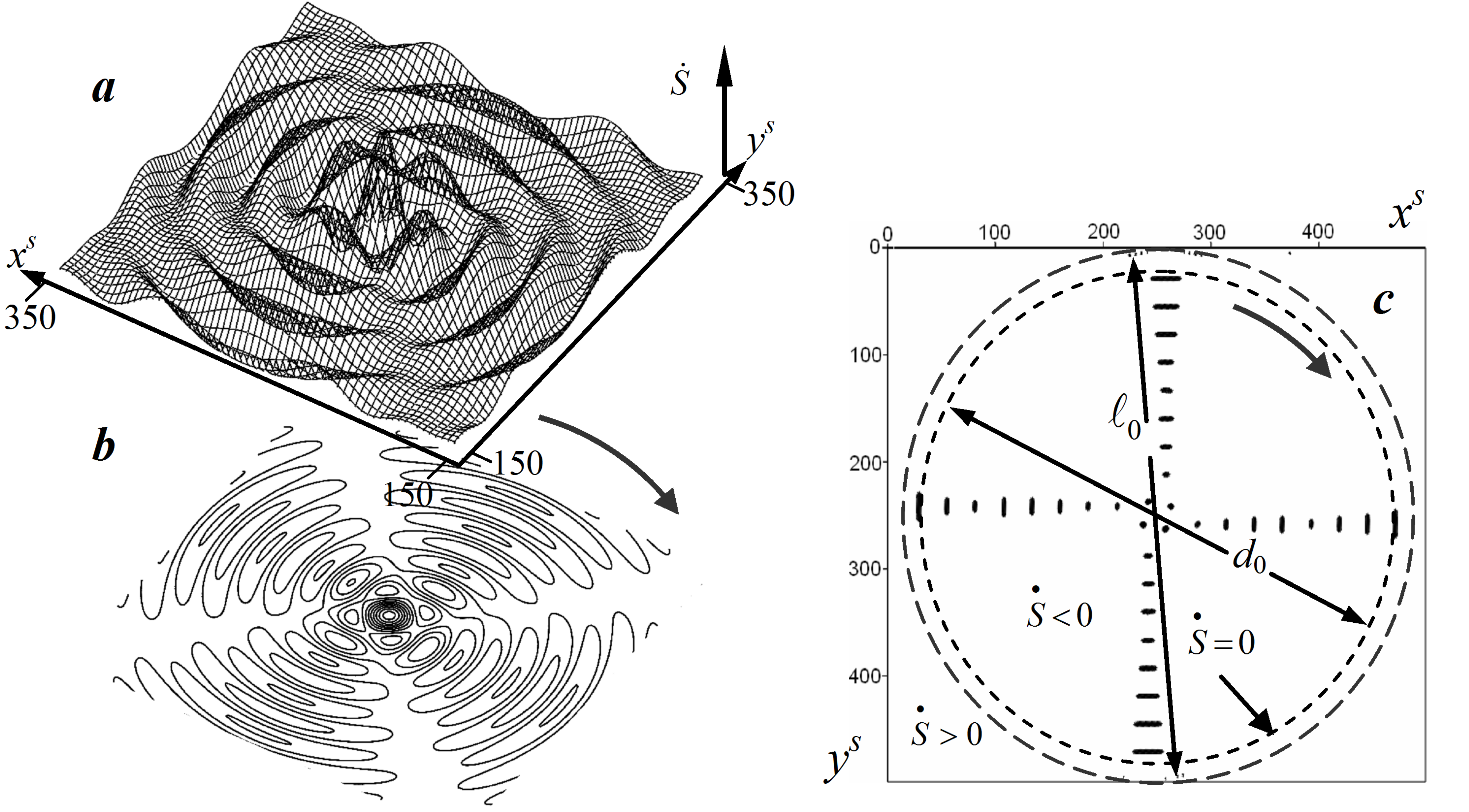}}
\caption{The total change in entropy in a plane layer of the tornado basin (a), its projection in the form of isolines (b) for the central region of the tornado that characterize the region of self-organization (c) and the boundary of the localized vortex structures for the calculation of the total area.}
\label{fig2tot}
\end{figure}

The sum of the external ($\sigma ^{e} $) and internal thermodynamic fluxes ($\sigma ^{i} $), is characterized by the full rate of change of entropy $\mathop{S}\limits^{\bullet } $. The decrease in entropy with time corresponding to processes of self-organization of vortex structures. The transition from the reduction of the entropy to its growth allows us to define the boundary of the vortex tornado basin, in which stable self-organized vortex systems exist. As follows from Fig. \ref{fig2tot}, the entropy of each tornado vortex layer decreases with time: $\mathop{S}\limits^{\bullet } <0$. This is achieved with a value of sources of momentum -- the reversible flows of entropy $\sigma ^{e} <0$, which are compared with the entropy production $\left|\sigma ^{e} \right|>\sigma ^{i} $. The condition of the limited lateral tornado is a value $\mathop{S}\limits^{\bullet } =0$, that is $\sigma ^{e} +\sigma ^{i} =0$ on the vortex-formation boundary (Fig.~\ref{fig2tot}c). Outside the vortex tornado basin entropy increases ($\mathop{S}\limits^{\bullet } >0$). The zone of self-organization decreases over time, which finally leads to the collapse of a tornado.

\subsection{Hydrodynamic conditions of self-organization in a tornado,}

established for the first time, follow from the thermodynamic conditions. Boundedness of tornado in height is owing to the absence of vortex formation due to the high viscosity. The transverse size of the tornado is caused by the competition between the rate of a velocity and a momentum distribution in terms of the viscosity. Positive feedback between the projections of horizontal velocities leads to sharpening. The intensity of the sources and sinks of movement causes the limitation on the modulus of the horizontal velocity: $\vartheta ^{*2} \ge 4q^{*} /\alpha _{1}^{*} (1-4c_{2}^{2} /9)$. Numerical calculations of such problem with the sharpening have shown the possible existence of enormous speeds of air movement directed to the nucleus. Also the approach allows us to calculate the spatial distribution of pressure in the tornado basin with change of height by the comparison with external pressure.

Restriction on the constant of the Kuramoto-Tsuzuki equation for the emergence of self-organization is a condition: $A_{2}^{2} \le \alpha _{2}^{*2} /\alpha _{1}^{*2} $. The left twisting condition in the topological charge is: $m=-1$, right -- $m=+1$. Restriction on the modulus of the topological charge for the formation of a tornado: $\left|m\right|\le 2$. The transition to the values of the topological charge $\left|m\right|>2$ firstly, leads to the huge cyclone model, i.e. atmospheric structures without a trunk with a significantly lower velocities and, secondly, to the emergence of "eye" -- the cyclone core.

The localized structure described and obtained in the framework of the sharpening problems on the basis of the thermodynamic approach is not unique. Similar effects occur in the thermal and diffusion processes, ferromagnetic materials. But in this article, self-organization in a regime with peaking for hydrodynamic problems was described by authors for the first time. This approach provides sufficiently uncontroversial physical results.

Thus, in this paper, the authors on the basis of the created model and the results of numerical calculations attempt to answer on all posed in the beginning of this article.

\bibliographystyle{my-h-elsevier}

\begin{thebibliography}{10}

\bibitem{Arsen}
S.A. Arsen'yev,
Mathematical modeling of tornadoes and squall storms,
GEOSCIENCE FRONTIERS {\bf 2}(2) (2011) 215$-$221.

\bibitem{Brooks}
E.M. Brooks,
The tornado-cyclone,
Weatherwise {\bf 2}(2) (1949) 32$-$33.

\bibitem{Brow}
K.A. Browning,
Airflow and precipitation trajectories within severe local storms which travel to the right of the winds,
J. Atmos. Sci. {\bf 21} (1964) 634$-$639.

\bibitem{DavJ}
R.P. Davies-Jones,
Streamwise vorticity: The origins of updraft rotation in supercell storms,
J. Atmos. Sci. {\bf 41} (1984) 2991-3006.

\bibitem{Flora}
S.D. Flora,
Tornadoes of the United States,
Oklahoma, (1953) 194 pp.

\bibitem{HF}
G.M. Heymsfield,
Kinematic and dynamic aspects of the Harrah tornadic storm analyzed from dual-Doppler radar data,
Mon. Wea. Rev. {\bf 106} (1978) 233$-$254.

\bibitem{Just}
A.A. Justice,
Seeing the inside of a tornado,
Monthly Weather Rev. {\bf 58} (1930) 57$-$58.

\bibitem{KlWil}
J.B. Klemp, R. Wilhelmson,
Simulations of right- and left-moving storms produced through storm splitting,
J. Atmos. Sci. {\bf 35} (1978) 1097$-$1110.

\bibitem{KKM}
S.P. Kurdyumov, E.S. Kurkina, G.G. Malinetskii,
Regimes with peaking. Achievements and Prospects. Problems of numerical analysis and applied mathematics, Dedicated to the anniversary of AA Samarskii,
Lviv. Ukraine. September 13-16 (2004).

\bibitem{KuTs}
Y. Kuramoto, T. Tsuzuki,
On the formation of dissipative structures in reaction-diffusion systems,
Progr. Theor. Phys. {\bf 54}  (1975) 687$-$699.

\bibitem{LemD}
L.R. Lemon, C.A. Doswell,
Severe thunderstorm evolution and mesocyclone structure as related to tornadogenesis,
Mon. Wea. Rev. {\bf 107}  (1979) 1184$-$1197.

\bibitem{Lane}
F.W. Lane,
The elements rage,
London. (1966) 279 pp.

\bibitem{Rasm}
E.N. Rasmussen, J. Straka, R. Davies-Jones, D.A. Doswell, F. Carr, M. Eilts, D.MacGorman,
Verification of the Origins of Rotation in Tornadoes Experiment: VORTEX,
Bull. Amer. Meteor. Soc. {\bf 75}  (1994) 997$-$1006.

\bibitem{SamM}
A.A. Samarskii,
Mathematical modeling and computer experiment,
Vestn. USSR {\bf 5} (1979) 38$-$49.

\bibitem{SKAM}
A.A. Samarskii, S.P. Kurdyumov, T.S. Akhromeeva, G.G. Malinetskii,
Modeling of nonlinear phenomena in modern science,
Computer Science and Technology progress.-Moscow: Science. (1987) 69$-$91.

\bibitem{SamC}
A.A. Samarskii,
Computers and Nonlinear Phenomena: Computers and modern science. Auto. foreword. Samarskii~A.A.,
Moscow: Science. (1988) 192 pp.

\bibitem{Schl}
R.E. Schlesinger,
A three-dimensional numerical model of an isolated thunderstorm. Preliminary results,
J. Atmos. Sci. {\bf 32} (1975) 835$-$850.

\bibitem{Weg}
A. Wegener,
Wind und Wasserhosen in Europa. In: Die Wissenschaft,
Bd. {\bf 60}, Braunschweig.  (1917) 301 pp.

\bibitem{WicW}
L.J. Wicker, R.B. Wilhelmson,
Simulation and analysis of tornado development and decay within a three-dimensional supercell thunderstorm,
J. Atmos. Sci. {\bf 52} (1995) 2675$-$2703.

\end{thebibliography}

\end{document}